\documentclass[sigconf]{acmart}
\copyrightyear{2026}
\acmYear{2026}
\setcopyright{cc}
\setcctype{by}
\acmConference[ICAIF '26]{7th ACM International Conference on AI in Finance}{November 14--17, 2026}{Milan, AA, Italy}
\acmBooktitle{7th ACM International Conference on AI in Finance (ICAIF '26), November 14--17, 2026, Milan, AA, Italy}
\acmDOI{10.1145/3858063.3858742}
\acmISBN{979-8-4007-3087-0/2026/11}

\usepackage{booktabs}
\title{What Prediction-Market Contracts Reveal at Launch: Trading Volume and Dispute Risk}

\author{Ahmed Mahrous}
\orcid{0000-0003-4694-5336}
\affiliation{%
  \institution{King Abdullah University of Science and Technology}
  \city{Thuwal}
  \country{Saudi Arabia}
}
\email{ahmed.mahrous@kaust.edu.sa}

\author{Roberto Di Pietro}
\orcid{0000-0003-1909-0336}
\affiliation{%
  \institution{King Abdullah University of Science and Technology}
  \city{Thuwal}
  \country{Saudi Arabia}
}
\email{roberto.dipietro@kaust.edu.sa}

\renewcommand{\shortauthors}{Mahrous and Di Pietro}

\begin{abstract}

We address the question: ``What does a prediction-market contract reveal at launch about its future?'' Using a near-census of 802{,}863 resolved markets listed on Polymarket, a leading prediction market, between January 2023 and May 2026, and only information available at listing (the market's title and description, named resolution source, and contract metadata), we predict lifetime trading volume and whether settlement is later disputed.

Contract information predicts trading volume well (out-of-sample $R^2 = 0.747$ for log volume), and most of the signal is interpretable: market topic provides the largest gain ($+0.289$ $R^2$ over a $0.220$ contract-feature baseline), wording and resolution-rule design add smaller but measurable improvements ($+0.027$ combined), and neural text models add little beyond the interpretable features. This performance does not survive a period of platform change coinciding with a shift in listing formats: accuracy in predicting whether a market reaches \$1{,}000 falls from AUC $0.940$ under same-period evaluation to $0.577$ when training on past quarters and testing on later ones. The loss concentrates in markets whose question forms were absent from training (76.9\% of test markets; on familiar forms the model still reaches $0.778$), and a decomposition attributes more of it to a changed relationship between contract characteristics and demand than to an outdated mapping from text to features.

Dispute predictors are more stable but less precise: launch-time features rank dispute risk far better than chance (AUC $0.78$), yet at a 0.31\% base rate precision is too low for case-by-case prediction (average precision $0.012$), so the scores suit ranking groups of markets rather than forecasting individual disputes.

The framework and results, besides their intrinsic interest, are actionable and likely extensible to other platforms, supporting further research in prediction markets.

\end{abstract}



\begin{document}

\maketitle

\section{Introduction}
\label{sec:intro}

Prediction markets have grown from small experimental venues into
large-scale platforms for trading on political, financial, crypto, sports, and other
real-world events; Polymarket's 2024 U.S.\ presidential-election market alone
generated more than \$3.6 billion in reported trading volume
\cite{tsang2026anatomy}. Most research studies these markets after trading
begins, asking whether prices aggregate information into accurate forecasts.
Platforms face an earlier design problem: newly listed contracts differ
greatly in what happens next, along two dimensions. Traction is extremely
uneven---in our data of more than 800{,}000 resolved Polymarket contracts,
the top 1\% of markets carry 62.4\% of all trading volume. Dispute risk is
rare but costly---0.31\% of markets whose dispute history is observable are
formally challenged, yet disputed markets carry 5.8\% of total volume across
the analysis panel (8.6\% within the dispute-observable subset). Both are set
in motion at listing, before order flow exists, when the
contract text and its terms are essentially the only market-specific
information available. 
We therefore ask: ``What does a
prediction-market contract reveal at launch about its later trading volume and
dispute risk?''

Every predictor is derived from contract information ordinarily set at
launch: the market's topic, its question and description text, and
characteristics of its resolution rules. Two features of the data shape the
design: related contracts (for example, all markets on one football match)
share attention and resolution logic, so we group them into events never
split across training and test; and because disputes are rare and
concentrated in particular subjects, we complement the predictive analysis
with a matched case-control comparison against non-disputed markets from the
same topic, platform quarter, and resolution-source type.

\noindentparagraph{Contributions}
We make three contributions.

\begin{figure}[!t]
  \centering
  \includegraphics[width=0.70\columnwidth,trim=0 1 0 11,clip]{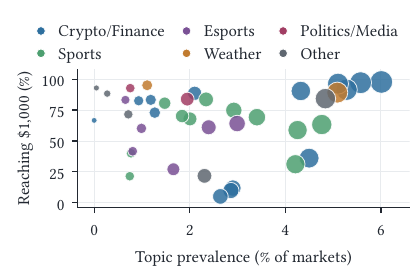}
  \Description{Scatter plot with one point per dominant NMF topic, plotting
  topic prevalence against the share of markets reaching one thousand
  dollars of volume; colors indicate the broad topic family (Crypto/Finance,
  Sports, Esports, Weather, Politics/Media, Other) and bubble area indicates
  the number of markets in the topic.}
  \caption{Each point is one NMF topic, sized by market count. Topic separates high- from low-traction segments.}
  \label{fig:topic-prevalence-traction}
\end{figure}

\noindentparagraph{Contract traction predictability}
We show that contract information predicts trading volume well
under same-period cross-validation, and most of this signal is interpretable. A
gradient-boosted tree model reaches out-of-sample $R^2=0.747$ for
log-transformed volume
and AUC $=0.940$ for whether lifetime volume reaches \$1{,}000. Knowing what
the market is about matters most: lexical topics add $0.289$ $R^2$ over a
basic contract-feature baseline of $0.220$ (Fig.~\ref{fig:ceiling-ladder}),
and topic alone separates high- from low-traction segments
(Fig.~\ref{fig:topic-prevalence-traction}). Conditional on topic, wording adds $0.021$
and rule drafting another $0.006$. Neural models add little, so interpretable
representations capture most of the launch-time signal available to the
tested models. The large remaining gain comes from interactions between
features rather than curvature in any one of them: for example, the same launch feature
can predict differently depending on the market's topic.

\noindentparagraph{Dispute predictability}
Launch-time features rank dispute risk
far better than chance (AUC $=0.78$), but at a 0.31\% base rate precision is
too low for individual prediction (average precision $0.012$), so we treat
dispute models as tools for grouping markets by risk. Adjusted analyses
identify which characteristics carry this risk: each additional grade level of
reading difficulty is associated with 23\% higher dispute odds; contracts
with no identifiable or successfully parsed resolution source have roughly
five times the adjusted odds of those using a mechanical feed; and an
anticipated-ambiguity pattern appears in edge-case drafting, where having a
clause is consistent with creators anticipating trouble while fuller coverage
among them is associated with lower odds. These are associations, not causal
estimates, but they identify contracts that may warrant review before launch.

\noindentparagraph{Temporal change and regime dependence}
Third, strong same-period prediction is not robust to
temporal change. The model with same-period AUC $0.940$ falls to $0.577$ when
trained on past quarters and evaluated on later ones. The failure is
concentrated in novel forms: 76.9\% of test markets use question
forms---wording templates that differ only in numbers and dates---never
observed in training, while on familiar forms the model still reaches AUC
$0.778$. Re-learning the text representation helps, but re-estimating the
feature--volume relationship recovers substantially more of the loss, so
launch-time volume models are regime-dependent and must be re-estimated as
the platform introduces new market forms.

\noindentparagraph{Roadmap}
Section~\ref{sec:related} places the study among work on prediction-market
quality, financial NLP, and temporal generalization.
Section~\ref{sec:data} describes the sample, target variables, launch-time features,
and the event-grouped evaluation methodology. Section~\ref{sec:traction} reports
the volume results and the feature ladder, and
Section~\ref{sec:robustness} checks them against template leakage and
alternative specifications. Section~\ref{sec:temporal} turns to
forward-in-time evaluation and isolates the role of novel question forms.
Section~\ref{sec:disputes} covers dispute prediction and the matched case-control risk factors. Section~\ref{sec:discussion} discusses deployment
implications, limitations, and future work.

\section{Related Work}
\label{sec:related}

\noindentparagraph{Prediction markets and market quality}
A prediction market turns a claim about an uncertain event into a tradable
security and ends by applying a resolution rule that maps the realized
outcome into a payoff. Most research studies the intervening stage: whether trading aggregates dispersed information into accurate
forecasts, and which conditions support that process
\cite{plott1988rational,wolfers2004prediction}.
Aggregation is possible but not automatic; it depends on the securities, the
trading mechanism \cite{hanson2007logarithmic}, and liquidity
\cite{chordia2008liquidity}.
Recent work carries these questions to modern platforms: the more liquid venue tends to
lead price discovery, with Polymarket leading Kalshi especially when its
activity and large-trade order flow are greater \cite{ng2026price}; and
within Polymarket's 2024 presidential market, arbitrage deviations and price
impact declined as participation expanded \cite{tsang2026anatomy}.
Informative prices are thus produced through participation, liquidity, and
informed traders rather than guaranteed. None of these studies asks how factors observable at
listing relate to subsequent market outcomes---the gap we fill.

\noindentparagraph{Contract language and financial NLP}
Financial text analysis shows that language observed before an outcome can
carry information about subsequent economic behavior
\cite{gentzkow2019text}, and that interpretable structure extracted from a
document predicts market outcomes \cite{wu2025press}. Crowdfunding is a close
parallel to our volume task: the language of a project description at launch
helps predict whether the campaign attracts funding
\cite{mitra2014language}. Prediction-market text, however, is not merely descriptive or persuasive; it also defines the security itself. Prior work on prediction-market contract design examines which uncertainties can be expressed as tradable claims \cite{adegbenro2026legibility}, how ambiguous event definitions block arbitrage across venues \cite{gebele2026semantic}, and whether resolution rules invite manipulation---settlement manipulation of Polymarket's five-minute Bitcoin contracts is absent from the fifteen-minute variant, making contract horizon a design lever \cite{dai2026settlement}. These questions are closely related to language because the contract's wording determines both the claim being traded and the conditions under which it settles. Unlike this prior work, we ask whether linguistic features observable at launch predict subsequent trading volume and dispute risk.

\noindentparagraph{Temporal generalization and concept drift}
Our temporal analysis relates to research on dataset shift, in which the joint distribution underlying the training data differs from that encountered at deployment \cite{moreno2012unifying}. In our setting, such shifts can arise from changes in the topics listed, the introduction of new contract formats, or changes in the trader population. Historical performance then need not carry over to deployment, a problem widely documented in language models and deployed financial forecasters \cite{lazaridou2021mind,jang2025regime}. We ask whether models built from contract-language features remain valid as Polymarket evolves, document a substantial regime change in which models that perform strongly within one period generalize poorly to later markets, and separate two explanations: \emph{text-representation drift}, in which the mapping from text to features has changed, and \emph{relationship drift}, in which the features still describe contracts adequately but the demand associated with them has moved.

\noindentparagraph{Prediction-market data resources}
Recent datasets expose different parts of the modern prediction-market
lifecycle---trade fills and oracle events \cite{jia2026unlocking},
contract rules paired with timestamped news and order books for evaluating
language models \cite{polybench2026}, and settlement legibility across venues
\cite{adegbenro2026legibility}---making trading, price formation, forecasting, and
settlement observable. Our dataset is complementary in what it measures:
alongside each contract's full text, it supplies wording and rule-design
variables---readability, edge-case coverage, conditional language, a
source-verifiability taxonomy---plus template and question-form groupings,
linked to lifetime volume and formal dispute status. Closest to our dispute
analysis, language models reproduce UMA's rulings once a dispute exists but
cannot say in advance which markets will be disputed \cite{wen2026disputes};
we reach a compatible conclusion from the opposite direction
(Sec.~\ref{sec:dispute-pred}).

\section{Data and Empirical Design}
\label{sec:data}

Prediction-market data pose four empirical challenges, each paired here with
a remedy: related contracts are statistically dependent (evaluation grouped
by event, Sec.~\ref{sec:models}); recurring title templates can inflate
measured generalization (template exclusion, Sec.~\ref{sec:templates}); the
platform changes rapidly (temporal cutoffs, Sec.~\ref{sec:temporal}); and
formal disputes are rare and concentrated in particular subjects (matched
case-control comparison, Sec.~\ref{sec:disputes}).

\subsection{Sample and Outcomes}
\label{sec:data-outcomes}

On Polymarket, each \emph{market} is a contract defined by a question, a
free-text description containing the resolution rules, and contract metadata
such as its scheduled close date, resolution settings, and referenced
resolution source. Related markets are grouped into
\emph{events}---all markets of one match, or all strikes of one price
ladder---that share a page, a comment stream, and often resolution logic;
some events also link mutually exclusive outcomes (e.g., the candidates in
one race) through the platform's \emph{neg-risk} mechanism, though each
member contract remains an ordinary yes/no market. When a market ends, a
proposed outcome is submitted to the UMA Optimistic Oracle and becomes final
unless someone challenges it during the challenge period
\cite{polymarket2026resolution}. A market is \emph{disputed} if its oracle lifecycle log
records at least one such challenge. That log exists for 65.9\% of the sample
(those without it are mostly earlier listings), and a dispute is invisible
without it, so all dispute analyses are restricted to this
\emph{dispute-observable} subsample.

We collect all market records from the platform's public API for
January 2022 through June 2026 and keep the 818{,}431 that closed with a
recorded resolution. Two exclusions follow from the design: 15{,}473 markets
with no valid scheduled end date (planned duration, a core predictor, is
undefined for them) and 95 markets whose categorical settings occur in fewer
than 50 markets---too few to estimate or match on (together 1.9\% of resolved
markets). The analysis sample holds 802{,}863 markets in 391{,}955 events
(Table~\ref{tab:data}), spanning January 2023 to May 2026. Its composition
shifts sharply over that window---74.3\% of the sample was created in 2026,
coinciding with a shift toward high-frequency recurring title templates such
as 15-minute cryptocurrency price windows---which is what the
temporal design of Sec.~\ref{sec:temporal} is built to test.

We study two outcome families. \emph{Trading volume} uses the platform's
recorded lifetime volume in dollars. Volume is extremely concentrated
(median \$3{,}542). We
model log-transformed lifetime volume, $y=\log(1+\mathrm{volume})$ (hereafter,
\emph{log-transformed volume}), and, as a binary outcome, whether a market
reaches \$1{,}000 in lifetime volume (68.5\% of markets do). Results are
insensitive to \$500 or \$5{,}000 thresholds
(Sec.~\ref{sec:addl-robustness}). \emph{Disputes}: 1{,}634 of the
529{,}408 dispute-observable markets (0.31\%, about one market in 320) had
their resolution formally challenged.

\begin{table}[!htbp]
  \caption{Analysis sample and target variables. The market is the unit of analysis;
  the event is the cluster used for all splits and standard errors.}
  \label{tab:data}
  \centering\footnotesize
  \renewcommand{\arraystretch}{0.90}
  \begin{tabular}{lr}
    \toprule
    \multicolumn{2}{l}{\textit{Panel}}\\
    Analysis sample (Sec.~\ref{sec:data-outcomes}) & 802{,}863 \\
    Events (clusters) & 391{,}955 \\
    Markets per event (median / mean / max) & 1 / 2.05 / 144 \\
    Creation window & 2023-01 -- 2026-05 \\
    Share created in 2025 / 2026 & 23.6\% / 74.3\% \\
    \midrule
    \multicolumn{2}{l}{\textit{Trading volume}}\\
    Volume: median / mean & \$3{,}542 / \$84{,}404 \\
    Volume Gini; top-1\% / top-10\% share & 0.927; 62.4\% / 88.6\% \\
    Markets with volume $\ge \$1{,}000$ & 68.5\% \\
    \midrule
    \multicolumn{2}{l}{\textit{Resolution}}\\
    Decisive binary settlement & 799{,}372 (99.6\%) \\
    50--50, void, or other settlement & 3{,}491 (0.4\%) \\
    Dispute-observable (oracle log recorded) & 529{,}408 (65.9\%) \\
    Disputed (0.31\% of dispute-observable) & 1{,}634 \\
    \bottomrule
  \end{tabular}
\end{table}

\subsection{Launch-Time Features}
\label{sec:representation}

Predictors are restricted to contract fields ordinarily set at listing and
the listing date (we use creation, launch, and listing interchangeably);
realized quantities such as prices, trader counts, and resolution outcomes
are excluded from every feature set. Four blocks are used.
\emph{Structured}: scheduled market duration, resolution settings,
resolution-source location and type, listing quarter, and market type (e.g.,
membership in a neg-risk group). \emph{Topic}, capturing what the market is
about: lexical topics from nonnegative matrix factorization (NMF
\cite{lee1999learning}, $k{=}40$) of TF--IDF-weighted question+description
text, plus $k$-means clusters ($k{=}40$) of e5-small sentence embeddings;
both are re-fit inside each training fold when used predictively. For
descriptive tables only, topic labels were proposed independently by two LLMs
from cluster keywords and representative markets under a fixed output schema,
then human-reviewed. \emph{Wording}, capturing how the question is phrased:
question and description lengths, quantitative anchors (prices, percentages,
dates, times), a specificity index (one point each for an explicit source,
date, and quantitative anchor, minus a penalty for vague resolution
phrases),
and a flag for such vague phrases (e.g., ``significant,'' ``credible
reporting''). \emph{Rule design}, capturing how the resolution rules are
drafted: edge-case coverage (how many of a fixed list of contingency
phrases---postponement, cancellation, ties, an unavailable source---the text
contains), conditional language (counts of markers such as ``if'' and
``unless''), the presence and length of a dedicated resolution paragraph,
Flesch--Kincaid readability, and a source-verifiability taxonomy based on the
type and public accessibility of the named source. All wording and
rule-design measures are deterministic functions of the contract
text---fixed keyword and regular-expression rules plus standard readability
formulas---and the released annotation guide gives their exact definitions.

\subsection{Models and Evaluation}
\label{sec:models}

We estimate a nested ladder of models, each adding one representation layer,
so comparisons show where predictive information resides: \textbf{M1} linear
regression (log-transformed volume) and $\ell_2$-regularized logistic regression (volume
$\ge$ \$1{,}000) on basic contract features (structured metadata, text
lengths, and numeric/date/time indicators); \textbf{M2} adds NMF topics;
\textbf{M3} adds embedding clusters instead; \textbf{M4} adds both;
\textbf{M5} is gradient-boosted trees (LightGBM \cite{ke2017lightgbm}) on the
M4 features, capturing nonlinearities and interactions the additive models
M1--M4 cannot. As a deep-text benchmark, \textbf{M6} replaces the
interpretable text layer with (a) frozen e5-small \cite{wang2022e5} or BGE-M3
\cite{chen2024bge} embeddings fed to the same boosted learner, and (b) an
e5-small encoder fine-tuned end to end with two output heads sharing one
encoder, one predicting log-transformed volume and one whether the market reaches
\$1{,}000. BGE-M3 is included to confirm the null result is not an artifact
of limited encoder capacity.\footnote{AdamW, one epoch, encoder learning rate
$2{\times}10^{-5}$, heads $10^{-3}$.} Hyperparameters are fixed in advance
(library defaults for the linear models; LightGBM: 600 trees, learning rate
0.05, 63 leaves, 80\% subsampling) and never tuned on evaluation data, so the
ladder compares representations rather than tuning effort. To isolate richer
wording and rule-design measures beyond M1, we also estimate a separate linear
block
ladder on the same folds---topics and era; duration and resolution settings;
wording; ambiguity; remaining rule-design
measures---from which the reported $+0.021$ and $+0.006$ increments are
paired out-of-fold changes, not steps in M1--M5.

Sibling markets within an event share text, trader attention, and resolution
logic, so a random split would place siblings of a test market in training
and overstate performance. Evaluation uses 5-fold cross-validation
grouped by event, with all text representations and models fit only on
training events, scored on the identical held-out events, and metrics pooled
over out-of-fold predictions. Regression is scored by out-of-sample $R^2$ on
log-transformed volume and classification by AUC---the probability that a randomly chosen
high-volume market is ranked above a low-volume one, so 0.5 is chance.
Grouped cross-validation still measures same-period skill, and recurring
title templates appear on both sides of fold boundaries;
Sec.~\ref{sec:robustness} therefore re-estimates everything after excluding
them, and Sec.~\ref{sec:temporal} imposes temporal cutoffs assigning whole
events by first-market creation time, so no future sibling informs a
past-trained model. For inference on performance gaps we bootstrap
whole events (200 replicates), keeping dependent siblings
together, and cluster standard errors at the event level; parametric AUC
comparisons overstate significance by treating markets as independent, so we
rely on the event bootstrap.

Disputes require a different design. Formal disputes are rare (0.31\%) and
concentrated by subject, quarter, and resolution source, so a plain
regression on the full sample devotes almost all of its fit to the handful of
subjects where disputes occur. Following standard epidemiological
practice for rare adverse events, Sec.~\ref{sec:disputes} instead compares
each disputed market with matched non-disputed controls from the same topic,
platform quarter, and source-type stratum, so that contract-design risk
factors are not confounded with what the market is about or when it was
listed.
\section{Predicting Trading Volume}
\label{sec:traction}

Under same-period cross-validation, a market's trading volume is highly predictable
from its contract alone. This section shows how predictable, where the signal
resides, and how little the tested neural models add.

\noindentparagraph{Launch-Time Predictability}
\label{sec:ladder}

\begin{figure}[!htbp]
  \centering
  \includegraphics[width=0.70\columnwidth,trim=0 1 0 1,clip]{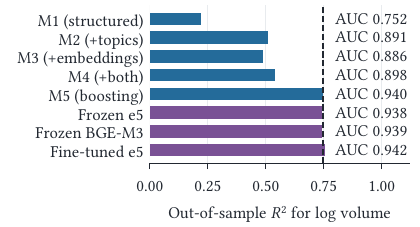}
  \Description{Model ladder showing out-of-sample R squared for
  log-transformed volume and AUC increasing from basic contract features to text
  models, with neural text benchmarks near the same ceiling.}
  \caption{Out-of-sample fit as feature blocks are added. Topic contributes
  the largest increment.}
  \label{fig:ceiling-ladder}
\end{figure}

\begin{figure}[!htbp]
  \centering
  \includegraphics[width=0.70\columnwidth,trim=0 1 0 2,clip]{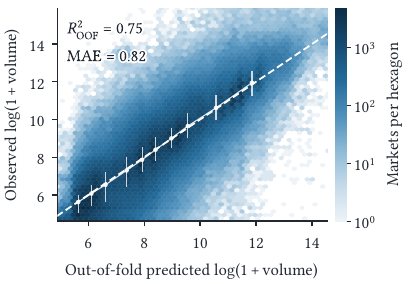}
  \Description{Hexagonal-binned density of out-of-fold predicted
  log-transformed volume on the horizontal axis against observed
  log-transformed volume on the vertical axis, shaded by the number of markets
  per hexagon on a logarithmic scale. The mass lies along the diagonal. Ten
  white dots mark decile means of observed volume with white interquartile
  bars, and a white dashed diagonal line marks perfect calibration.}
  \caption{Out-of-fold predicted vs.\ observed volume (M5). Dashed line:
  perfect calibration; dots and bars: decile means and interquartile ranges.}
  \label{fig:predicted-observed}
\end{figure}

The boosted topic model (M5) reaches $R^2=0.747$ and AUC $=0.940$ out of
sample; the ladder in Fig.~\ref{fig:ceiling-ladder} shows where that comes from. The
basic contract-feature baseline (M1) explains $R^2=0.220$ at AUC $0.752$, and its
largest coefficients are consistent with greater volume for more precisely
specified contracts: an explicit resolution source has an estimated coefficient
of $+1.8$ to $+2.2$ for log-transformed volume, concrete quantitative anchors
(prices, dates, percentages) have coefficients of $+0.7$ to $+1.5$, and each
unit increase in log-transformed scheduled duration has a coefficient of
$+0.59$. Text then enters in two steps: lexical topics nearly double
explained variance (M1$\to$M2, $+0.289$ $R^2$ [$+0.286,+0.292$]), and
allowing nonlinearity and interactions over the same features adds another
large block (M4$\to$M5, $+0.207$ [$+0.204,+0.210$]). These gains are stable
across folds (standard deviation $\le0.012$) and every pairwise gap is
significant under the event-cluster bootstrap. The boosted model is also well
calibrated out of sample: 6.2\% of markets in the bottom decile of predicted
probability reach \$1{,}000 versus 6.3\% predicted, and 99.93\% versus
99.87\% in the top decile. The same holds for the volume regression
(Fig.~\ref{fig:predicted-observed}). Within a stable regime its scores could
therefore rank incoming listings directly; Sec.~\ref{sec:temporal} tests
whether that transfers to future ones.

\noindentparagraph{Market Topic}
\label{sec:subject}

Table~\ref{tab:topics} shows why the topic layer is so important: recurring
templates differ substantially in outcomes, a spread larger than the
variation associated with any single wording or rule-design feature we
measure. What a market is about, inferred from contract text alone, is thus
the strongest predictor of its trading volume, and the pattern holds across
all topics, not just these illustrative templates
(Fig.~\ref{fig:topic-prevalence-traction}).

\begin{table}[!htbp]
  \caption{Selected recurring title templates. Counts and volume shares use all
  markets; dispute rates use the dispute-observable subsample.}
  \label{tab:topics}
  \centering\footnotesize
  \renewcommand{\arraystretch}{0.90}
  \begin{tabular}{p{0.48\columnwidth}rrr}
    \toprule
    Example recurring title template & Markets & \shortstack{Volume\\$\ge$ \$1{,}000} & \shortstack{Dispute\\rate} \\
    \midrule
    Crypto up/down, 15-min windows & 343{,}098 & 70.5\% & 0.40\% \\
    Sports over/under totals & 67{,}149 & 55.3\% & 0.12\% \\
    Daily temperature ranges & 60{,}260 & 88.2\% & 0.39\% \\
    Tennis set-game totals & 11{,}448 & 24.1\% & 0.07\% \\
    Trump/political-media mentions & 10{,}998 & 86.0\% & 1.00\% \\
    \bottomrule
  \end{tabular}
\end{table}

\noindentparagraph{Neural Text Benchmarks}
\label{sec:neural}

Could a neural text model extract more from the same contracts? Two
comparisons indicate that it cannot, at least for the models tested. First,
interpretable lexical topics outperform embedding clusters (M3 is \emph{below} M2,
$\Delta R^2=-0.021$), and the two are only modestly complementary
($+0.031$). Second, the neural benchmark leaves little room above the
interpretable ceiling: raw frozen embeddings fed to the same boosted learner
match M5 to three decimals (0.743/0.744 vs.\ 0.747), the larger BGE-M3
encoder adds nothing over e5-small, and end-to-end fine-tuning ends only
slightly above ($R^2$ 0.754 vs.\ 0.747, $+0.007$ [$+0.005,+0.009$] under the
paired event-cluster bootstrap).\footnote{BGE-M3 encoded text truncated to
1{,}600 characters versus e5's 512-token window; given the null encoder
effect this difference is immaterial.} The 40 NMF topic features plus
structured features thus capture nearly all of the launch-time textual signal
the tested models can extract.

\noindentparagraph{Wording and Rule Design}
\label{sec:wording}

\begin{figure}[!htbp]
  \centering
  \includegraphics[width=0.78\columnwidth,trim=0 1 0 1,clip]{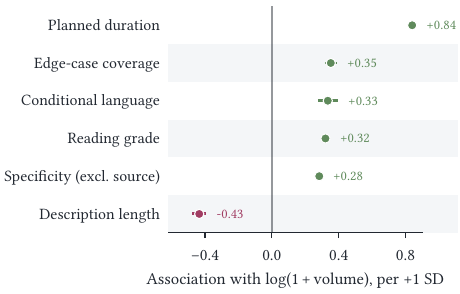}
  \Description{Forest plot of six continuous contract-design factors ordered by
  effect size, each shown as a point estimate with a 95 percent confidence
  interval, all on a one-standard-deviation contrast. Planned duration has much
  the largest positive estimate; edge-case coverage, conditional language,
  reading grade and specificity form a narrow positive band; description length
  is negative.}
  \caption{Contract-design factors with topic and launch-quarter fixed
  effects on a one-standard-deviation scale.}
  \label{fig:design-factors}
\end{figure}

Conditional on topic, how a contract is written still matters, but
marginally: the wording block adds $+0.021$ $R^2$ (positive in all folds) and
the rule-design measures---edge-case coverage, conditional language,
resolution-paragraph structure, readability, source verifiability---add a
further $+0.006$, about one-fiftieth of the topic contribution. The
direction is nonetheless informative: the specificity index rises
monotonically with contract success, from 50.5\% of markets scoring 0
reaching \$1{,}000 to 83.0\% at a score of 3. Precise rules attract more
trading, an effect large enough to order markets within a topic but too small
to overcome the topic effect itself.
Figure~\ref{fig:design-factors} compares these associations on a one-standard-deviation scale. Planned duration has by far the largest
positive association with log-volume ($+0.84$), while specificity, reading
grade, conditional language, and edge-case coverage form a narrower positive
band ($+0.28$ to $+0.35$); description length is the only negative factor
($-0.43$).

The boosted model's advantage over the linear ladder comes largely from how
launch features \emph{combine}, not from curvature in any one of them.
Constraining M5 to depth-one trees (single-feature nonlinearities allowed,
interactions not) closes only 25\% of the M4$\to$M5 $R^2$ gap; allowing
pairwise interactions closes 57\%; and a linear model with named interactions
(duration $\times$ topic, word counts $\times$ topic, duration $\times$ era)
closes 32\%---more than the additive-tree ceiling. The largest contrast is
duration $\times$ topic: the estimated duration slope is $+2.92$ for Bitcoin
Up/Down Micro-Markets, compared with $-0.03$ for NMF topic 26 (Basketball
Game Lines and Totals) and $+0.24$ for Social Media Post Count Markets.
Relative to the reference Crypto Up/Down Micro-Markets topic, whose slope is
$+2.09$, the latter interaction differences are $-2.12$ and $-1.85$,
respectively (both $p<0.01$). No launch feature has a
universal effect, so screening rules built on these features should be
topic-aware.

\section{Robustness Checks}
\label{sec:robustness}

The within-period results of Sec.~\ref{sec:traction} raise two concerns: the
model may be exploiting recurring title templates rather than generalizable
contract signal, and the findings may be sensitive to analytical choices.
Neither holds.

\noindentparagraph{Recurring Title Template Exclusion}
\label{sec:templates}

We apply an outcome-blind definition of recurring title templates: lowercase
each question, replace month and weekday names, money amounts, percentages,
and numbers with placeholders, and retain only forms occurring exactly once
in the full panel. This removes 666{,}967 markets (83.1\%) and leaves
135{,}896, on which folds, topics, and embedding clusters are re-derived. The
qualitative hierarchy holds: M1 reaches $R^2=0.230$/AUC $=0.729$, lexical
topics add $+0.154$ $R^2$, embedding and lexical topic models are nearly
tied, and interactions add $+0.225$, yielding M5 $R^2=0.652$/AUC $=0.891$.
The main result therefore does not require repeated question forms, although
the lower performance shows that repetition is part of the platform's
predictability.

\noindentparagraph{Additional Robustness}
\label{sec:addl-robustness}

Varying the volume threshold, the treatment of outliers, the encoding of
platform era, the unit of analysis, and the number of topics each moves the
reference specification by less than 0.01 in almost every row. Three
comparisons are substantively informative. Hand-coded regex wording proxies
\emph{without} learned topics fall back to the basic contract-feature baseline
(0.227/0.756), so simple feature engineering does not recover the topic
layer's contribution. Aggregating to the event level preserves strong
regression performance ($R^2=0.674$ vs.\ 0.747) while slightly improving
classification (AUC 0.956 vs.\ 0.940), so market-level clustering is not
producing the result. And the headline performance is not carried by any
single topic (Table~\ref{tab:drop_family}): crypto/finance---45\% of the
panel---produces the largest change, and sports removal raises $R^2$ to
0.785.

\begin{table}[!htbp]
  \caption{Leave-one-family-out robustness: each row recomputes the reference
  model's out-of-sample fit with one broad topic family excluded.}
  \label{tab:drop_family}
  \centering\footnotesize
  \renewcommand{\arraystretch}{0.90}
  \begin{tabular}{lrrr}
    \toprule
    Excluded family & \% panel & $R^2$ & AUC \\
    \midrule
    None (full model) & --- & 0.747 & 0.940 \\
    \midrule
    Crypto/Finance & 45.4 & 0.668 & 0.896 \\
    Sports & 28.0 & 0.785 & 0.957 \\
    Esports & 9.7 & 0.752 & 0.945 \\
    Other events & 8.1 & 0.755 & 0.941 \\
    Weather & 6.3 & 0.754 & 0.940 \\
    Politics/Media & 2.5 & 0.748 & 0.940 \\
    \bottomrule
  \end{tabular}
\end{table}

One finding warrants caution. Comparing sibling markets within
events looks attractive---siblings share event-level demand, so differencing
them seems to isolate drafting---but on this platform sibling wording often
encodes a market's \emph{role} within its event rather than a drafting
choice. A within-event regression produced an implausible coefficient of
$+6.29$ for log-transformed volume on the vague-language flag; an audit traced
it to composition,
the flag marking boilerplate that appears almost exclusively on the flagship,
highest-volume market of its event. Restricting the comparison to
structurally identical siblings shrinks the coefficient to $+0.21$
($p=0.29$). Within-group designs on prediction-market panels can thus mistake
a market's role for its wording, and should be audited accordingly.

\section{Temporal Generalization}
\label{sec:temporal}

These results measure accuracy within the model's training regime. Prospective evaluation---training on the past and predicting the
future---asks whether those relationships survive the platform's own change.
They largely do not (Table~\ref{tab:temporal_summary}); this section
diagnoses why.

\begin{table}[!htbp]
  \caption{Predicting whether a market reaches \$1{,}000 (AUC; chance $=0.5$),
  boosted model. Recalibration is fit on the earliest 10\% of the test
  period.}
  \label{tab:temporal_summary}
  \centering\footnotesize
  \renewcommand{\arraystretch}{0.90}
  \begin{tabular}{@{}lr@{}}
    \toprule
    Evaluation setting & AUC \\
    \midrule
    Train and test in the same period (cross-validation) & 0.940 \\
    Train on the past, test on the future (cutoff 2025Q4) & 0.577 \\
    \quad question form never seen in training (77\% of test) & 0.575 \\
    \quad question form seen in training (23\% of test) & 0.778 \\
    \addlinespace
    Diagnostics on later-period data: & \\
    \quad before / after monotone rescaling (same 90\% subset) & 0.597 / 0.597 \\
    \quad later-period refit (hindsight diagnostic) & 0.940 \\
    \bottomrule
  \end{tabular}
\end{table}

\noindentparagraph{Platform Regime Change}
\label{sec:prospective}

Prospective evaluation trains on all events created up to a cutoff and tests
on everything after. Models achieving $R^2$ of 0.5--0.75 in same-period
evaluation perform far worse across regimes: cutoffs at 2025Q2 or 2025Q4
yield negative $R^2$ and AUC of 0.52--0.58---at or near chance---while only
the 2026Q1 cutoff, whose training window already contains the modern regime,
recovers usable performance ($R^2=0.627$, AUC $=0.906$). The drop from 0.940
to 0.577 coincides with the 2026 expansion of template micro-markets. Of the
panel, 74.3\% was created in 2026 (Table~\ref{tab:data}), when many prevalent
question forms were absent from earlier training windows. Whether change of this magnitude
recurs is open---it may prove a one-time transition---but a deployed model
cannot know that in advance, which motivates the diagnostics below.

\noindentparagraph{Structural Novelty}
\label{sec:novelty}

The prospective failure is concentrated where the platform introduces new
market forms. Two questions share a \emph{question form} if they match after
masking numbers, dollar amounts, percentages, and dates (so all ``Bitcoin
above \$X on date D?'' markets share one form). At the 2025Q4 cutoff, 76.9\%
of test markets have a form never seen in training; on those the past-trained
model has near-zero rank discrimination (AUC 0.487--0.575), while on familiar
forms it retains some (AUC 0.68--0.78). Markets outside the labeled recurring
families transfer better (AUC 0.674 vs.\ 0.516 pooled for the linear model)
but still fall short of their in-period ceiling of 0.803, so the
novel-template influx heightens the failure without being its only source.
Novel forms are not a marginal problem: conditional on topic and era,
first-of-kind markets (no prior near-duplicate at embedding cosine $\ge0.95$;
3.5\% of the panel) have an estimated coefficient of $+1.35$ for
log-transformed volume ($p<0.01$) and
$+1.25$ log-odds of reaching \$1{,}000, and volume declines with the number
of prior near-duplicates---so the markets the model handles worst are
disproportionately the high-volume ones.

\noindentparagraph{Representation vs.\ Relationship Drift}
\label{sec:decomp}

\begin{table}[!htbp]
  \caption{Attributing the 2025Q4 loss: out-of-sample $R^2$ on the same later
  markets, with each part kept from the past or re-learned from recent data.}
  \label{tab:drift_decomp}
  \centering\footnotesize
  \renewcommand{\arraystretch}{0.90}
  \begin{tabular}{lcc}
    \toprule
     & \multicolumn{2}{c}{Relationship (features$\to$volume)} \\
    \cmidrule(lr){2-3}
    Text representation & from the past & re-learned \\
    \midrule
    From the past & 0.011 & 0.772 \\
    Re-learned    & 0.249 & 0.773 \\
    \bottomrule
  \end{tabular}
\end{table}

Why does a model trained on the past fail on the future? Its knowledge has
two parts: the \emph{text representation}---how it reads contract text into
features---and the \emph{relationship} linking those features to volume.
Either could be outdated: the platform's language might have changed, or the
same kind of contract might now attract different demand. We test both by
re-learning each part from recent data, one at a time, and re-scoring the same
later markets (Table~\ref{tab:drift_decomp}). Re-learning the representation
recovers meaningfully ($R^2$ 0.011 to 0.249), but re-learning the
relationship recovers much more (0.011 to 0.772), and once the relationship
is re-estimated the representation adds almost nothing (0.772 vs.\ 0.773).

That also explains why a simpler correction fails. Shifting all predictions up
or down by a single constant would work if the model still ranked contracts
correctly and only its overall level were off---as it would be if demand had
risen or fallen across the board. But a uniform shift cannot change which
contracts the model ranks above which, and that ranking is what broke (AUC is
unchanged before and after the adjustment). Demand did not simply move up or
down for every market; the kinds of contracts attracting the most trading
changed relative to one another.

The practical fix is to retrain, not recalibrate. Retraining once per quarter
on recent data improves performance even through the transition: AUC
0.64--0.71 during the turbulent 2025 shift (higher on familiar question
forms, near-chance on novel ones), rising to 0.907 by 2026Q2 once the new
regime is represented in training. Using only the two most recent quarters
works about as well as using the entire history, so old data neither clearly
helps nor hurts. A deployed launch-time screen therefore needs three rules:
retrain at least every quarter, or reweight training data toward the current
regime \cite{jang2025regime}; track the share of never-before-seen question
forms as an early warning that predictions may not transfer; and treat
predictions on those forms as low-confidence until similar markets enter the
training data.

\section{Disputes: Prediction and Risk Factors}
\label{sec:disputes}

Disputes pose the same problem as rare adverse events in epidemiology, and we
borrow its logic: rarity blocks case-by-case prediction, but risk factors can
still be measured by comparing affected cases with matched unaffected
controls. This maps onto two questions---the \emph{predictive} one, whether
launch-time information forecasts which individual markets will be disputed
(it does, but only weakly), and the \emph{descriptive} one, which design
characteristics disputed contracts share (three stable patterns).

\noindentparagraph{Dispute Predictability}
\label{sec:dispute-pred}

Disputes are rare---1{,}634 of 529{,}408 dispute-observable markets
(0.31\%)---and launch-time prediction is correspondingly limited. A logistic
model reweighted for class imbalance ranks risk far better than chance
(out-of-sample AUC 0.78 under event-grouped folds); the more flexible
boosted model overfits the roughly 1{,}600 positives and does worse, at
0.561. AUC scores \emph{ordering} and is base-rate invariant---0.78 means a
randomly chosen disputed market outscores an undisputed one 78\% of the
time---while precision is not, so at 0.31\% even a well-ordered list is
overwhelmingly false positives. Average precision is
0.012 against a random baseline of 0.0031, and
reviewing the riskiest 5\% of markets raises the observed dispute rate to
approximately 1.7\% while capturing just 27.4\% of disputes. Launch text thus
supports \emph{triage}---concentrating review where risk is highest---but not
individual prediction, so we treat disputes as an associational outcome
rather than a predictive target. The descriptive
gradients are nonetheless steep. By resolution-source type, dispute rates
span a factor of nearly fifty: 0.08\% for markets resolved from an on-chain
oracle price feed (3{,}907 markets), 0.15\% from a sports official
(212{,}163), 0.26\% from a financial-market data feed (13{,}128), 0.66\% from
the description alone (126{,}708), 1.01\% where no source is identifiable
(12{,}746), and 3.79\%---more than $12\times$ the base rate---for
entertainment-industry references such as box-office trackers, though that
last rate rests on 18 disputes in 475 markets. The pattern is intuitive:
sources that report a numeric value directly are disputed least, while those
that require interpreting media coverage are disputed most, including markets resolved by
whether a public figure or outlet mentions a keyword (1.64\%).

\noindentparagraph{Matched Risk-Factor Analysis}
\label{sec:dispute-factors}

\begin{figure}[!htbp]
  \centering
  \includegraphics[width=0.78\columnwidth,trim=0 1 0 1,clip]{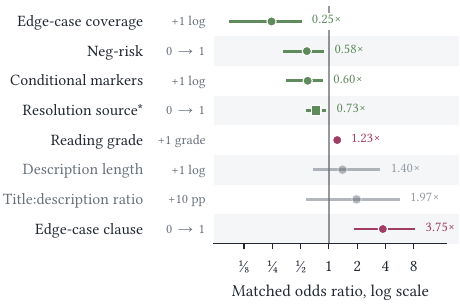}
  \Description{Forest plot of matched dispute odds ratios for resolution-source
  presence, neg-risk structure, edge-case coverage, conditional markers,
  readability, description length, title-to-description ratio, and edge-case
  clauses.}
  \caption{Contract characteristics of disputed markets: matched
  conditional-logit odds ratios with 95\% event-cluster bootstrap intervals;
  grey marks intervals crossing 1. Unstarred rows: topic-by-quarter-by-source-type
  match; starred row: topic-by-quarter match, Wald interval.}
  \label{fig:dispute-forest}
\end{figure}

Raw gradients confound contract design with subject, launch date, and source
type. In the primary design each disputed market receives up to four controls
from the same dominant NMF topic $\times$ platform quarter $\times$
source-type stratum, yielding 1{,}634 cases and 6{,}520 controls; five sparse
cells exhaust their available controls, and the conditional logit uses
8{,}154 observations in 225 informative strata with event-cluster bootstrap
CIs (Fig.~\ref{fig:dispute-forest}). Because this design holds source type
constant, it estimates only non-source characteristics; source presence,
type, and accessibility are estimated separately in models adjusting for
topic and quarter---a topic-by-quarter conditional logit for presence and a
full-sample Firth penalized logit \cite{firth1993bias} for type and
accessibility.

Three patterns survive every robustness variant we ran (keeping one disputed
market per event, splitting by platform era, excluding labeled template
families). \emph{Readability}: each additional Flesch--Kincaid grade level of
the rules raises dispute odds by 23\% [14\%, 34\%], the most stable
coefficient in the study. \emph{Verifiability}: the topic-by-quarter
source-presence model gives an odds ratio of 0.73 [0.57, 0.94]; in the Firth
model, no identifiable or successfully parsed source versus a mechanical feed
has odds ratio 4.93 [3.25, 7.47], and relative to no identifiable or
successfully parsed source, freely accessible public sources and paywalled or
otherwise restricted sources have odds ratios 0.56 [0.45, 0.68] and 0.18 [0.07, 0.44]
(Wald intervals). \emph{Edge-case drafting}: having any edge-case clause is
associated with $3.8\times$ higher dispute odds, while conditional on having
one, denser coverage is associated with lower odds (odds ratio 0.25). We read
this as anticipated ambiguity---creators add edge-case language when they
foresee trouble, so the clause marks risky contracts, and among those more
complete coverage accompanies fewer disputes. All of these characterize which
\emph{contract designs} end up in dispute, useful for prioritizing review, rather than causal effects of drafting.

\section{Discussion}
\label{sec:discussion}

\noindentparagraph{Implications for deployment}
Launch-time contract information is
useful but unevenly distributed, and the two outcomes it speaks to belong in
different workflows. A volume model can rank incoming listings directly, but
only inside a stable regime, so deployment requires the refit-and-monitor
discipline of Sec.~\ref{sec:decomp}. A dispute model instead
belongs in human review rather than an automated accept/reject rule, because
at a 0.31\% base rate even a well-ranked risk list is mostly false positives.
Resampling, cost-sensitive thresholds, and our class reweighting did not raise positive predictive value enough to change that.

\noindentparagraph{Scope, limitations, and future work} Our neural comparison covers two widely used open-weight sentence encoders---e5-small (33M parameters) and BGE-M3 (568M), evaluated frozen and, for e5-small, after in-domain fine-tuning. We therefore do not claim that no text model could improve on the reported results. Rather, returns to additional capacity are flat within this comparison: BGE-M3 adds nothing over the substantially smaller encoder, while fine-tuning on roughly 800{,}000 contracts improves $R^2$ by only $0.007$. Our conclusion is thus limited to the tested model class: interpretable features capture nearly all of the launch-time textual signal these models recover.

First, the analysis covers a single platform. Polymarket is large and heterogeneous, and our data span substantial changes within the platform, but its listing process, trader population, and mechanisms may differ from those of more curated or regulated venues. Cross-platform analysis could test which predictors transfer to venues such as Kalshi.

Second, the estimated relationships between contract design and outcomes are also associational, not causal. Although the matched design adjusts for topic, period, and source type, creators choose the remaining contract characteristics jointly, and the within-event audit (Sec.~\ref{sec:addl-robustness}) shows how an apparent wording effect can instead reflect a market's role or subject. Stronger causal evidence could come from field experiments that vary contract features. 

Finally, the temporal results identify structural novelty as a consequential deployment limitation: models trained on earlier contracts perform poorly on previously unseen question forms. Future work could develop novelty-aware systems that detect regime changes, report greater uncertainty on unseen forms, and update feature--outcome relationships online. 
\section{Conclusion}
\label{sec:conclusion}
In this paper, we investigated what a prediction-market contract reveals at launch, before any
trading occurred, about its later trading volume and dispute risk. We have adopted a data-driven approach, using a near-census of 802{,}863
resolved markets listed on Polymarket between January 2023 and May 2026.

Results are revealing and actionable:
launch-time text
predicts volume strongly and interpretably: what a market is about carries
most of the signal, wording and rule design add smaller topic-conditional
increments, and our handcrafted features match the neural encoders we tested.
But features do not act uniformly across topics, and the predictability is
regime-dependent: strong within a period, fragile across the platform's own
change, with the failure concentrated in new market forms and attributable
more to changed feature--demand relationships than to an outdated mapping from
text to features.
For disputes, launch text supports ranking markets by risk but not predicting
individual ones, and adjusted analyses identify three stable design
correlates: readability, source verifiability, and edge-case drafting. 

These findings identify what launch-time information can
reveal---which listings attract volume, which designs are associated with disputes---and
what it cannot: individual disputes, and volume once the regime shifts.
They also highlight methodological challenges specific to research on
prediction markets: dependence among related contracts within events,
repeated market templates, temporal regime change, and the evaluation of
extremely rare disputes.
\noindentparagraph{Data and code availability}
The data and code for reproducing the paper's figures and tables are available
at \url{https://doi.org/10.5281/zenodo.23376383}.

\bibliographystyle{ACM-Reference-Format}
\bibliography{references}

\end{document}